\documentclass{ifacconf}

\usepackage{graphicx}      
\usepackage{natbib}        
\usepackage{amsmath}
\usepackage{amssymb}
\usepackage{dutchcal}
\begin{document}
\begin{frontmatter}

\title{Robust reset control design for piezo-actuated nano-positioner in presence of hysteresis nonlinearity\thanksref{footnoteinfo}} 

\thanks[footnoteinfo]{This work was financed by Physik Instrumente (PI) SE \& Co. KG and co-financed by Holland High Tech with PPS Project supplement for research and development in the field of High Tech Systems and Materials.}

\author[First]{A. Sebghati} 
\author[First]{S. H. HosseinNia}

\address[First]{Department of Precision and Microsystems Engineering, Faculty of Mechanical Engineering, Delft University of Technology (e-mail: a.sebghati@tudelft.nl; s.h.hosseinniakani@tudelft.nl)}

\begin{abstract}                
In this paper, a robust nonlinear control scheme is designed for the motion control of a class of piezo-actuated nano-positioning systems using frequency-domain analysis. The hysteresis, the nonlinearity in the piezoelectric material, degrades the precision in tracking references with high frequency contents and different travel ranges. The hysteresis compensation by the inverse model, as the state-of-the-art solution, is not reliable alone. Therefore, a control framework with robustness against the remaining nonlinearity is needed. It is shown that there is an unavoidable limitation in robust linear control design to improve the performance. A robust control methodology based on a complex-order element is established to relax the limitation. Then, a constant-in-gain-lead-in-phase (CgLp) reset controller is utilized to realize the complex-order control. The control design is based on the sinusoidal input describing function (SIDF) and the higher-order SIDF (HOSIDF) tools. A constrained optimization problem is provided to tune the control parameters. The achieved improvements by the CgLp control is validated by the simulation.
\end{abstract}


\end{frontmatter}

\section{Introduction}
Piezoelectric-actuated nano-positioners play a vital role in the operation of the high-tech machines where ultra-precision motions are required. The ever-growing demands for a higher performance necessitate the piezo-positioners to precisely track high-speed periodic trajectories with a wide range of amplitudes. Therefore, there is a need to develop a motion control system to provide a wide bandwidth (BW), and a high and low open-loop gain below and above the BW frequency, respectively, for the full operational range (see e.g. \cite{NanopositioningTechs_Book_Piezo_Hysteresis}).

Unfortunately, there are sources of nonlinearity in a piezo-positioner, which makes it challenging to develop the desired motion control. Hysteresis is one of the significant sources that appears as an uncertainty in the DC gain of the piezo-positioners. An inverse model of the hysteresis is used to compensate for the nonlinearity as the state-of-the-art solution (see e.g. \cite{NanopositioningTechs_Book_Piezo_Hysteresis}). A full compensation is not practical due to the model complexity, rendering a need to robustness against the remaining uncertainty by motion control.

The currently used motion control schemes in industry are based on the linear control synthesis in the frequency domain. Among the linear control techniques, iso-damping control (\cite{IsoDamping_Car_Bode,IsoDamping_DataDriven_New}) makes the closed-loop system robust against the DC gain uncertainty by adding phase lead to the open loop. However, any linear lead operator around the BW frequency degrades the performance of the piezo-systems from the open-loop gain perspective. Therefore, a reduction in BW, and consequently in precision, is needed to have a robust linear motion control. The high trade-off in improving the different aspects of performance is due to the waterbed effect.

Reset control, introduced by \cite{clegg1958nonlinear}, has earned a growing attention in the high-tech industry. Although it is a nonlinear control framework, sinusoidal input describing function (SIDF) and higher-order SIDF (HOSIDF) allow for reset control design in the frequency domain (\cite{zhang2024_SIDF_HOSIDF_ResetControl}). The aforementioned trade-off is flexed in reset control design. Constant-in-gain-lead-in-phase (CgLp) reset control has been recently developed and utilized for the precision motion systems (\cite{saikumar2019CgLp}). It adds phase lead to the open loop while leaving the open-loop gain almost unchanged. Therefore, the CgLp element is potential to make a robust motion control with a higher BW and precision compared to linear motion control.

In this work, a model for a piezo-actuated positioner with the hysteresis nonlinearity is provided in Section \ref{sec: problem definition}. Then, a robust control scheme is presented for linear systems with the DC gain uncertainty in Section \ref{sec: control methodology}. In Section \ref{sec: robust control design}, a robust motion control is synthesized by a set of linear control elements. The problem is discussed over the result of the designed linear motion control. Next, a CgLp element is integrated into the whole control system to increase the precision by increasing the robustness at higher frequencies. The presented methodology for the CgLp control design does not exist in the literature based on the best knowledge of the authors. The parameters of the CgLp element are tuned by a constrained optimization problem. The designed nonlinear motion control is validated in the frequency domain by the HOSIDF tool and in the time domain in Section \ref{sec: simulation results}. It is shown how much trade-off between the robustness and the BW is relaxed by the simulations.\\
\textbf{Notations:} $\omega_.^{..}=2{\pi}f_.^{..}$, $\mathcal{w}_.^{..}=2{\pi}\mathcal{f}_.^{..}$; $\mathbb{R}_{+}$: $\mathbb{R}_{{\geq}0}$, $\mathbb{R}_{++}$: $\mathbb{R}_{>0}$
\section{Problem definition}\label{sec: problem definition}
\subsection{System description}\label{subsec: system description}
The one-directional stage P-621.1CD PIHera equipped with a piezoelectric actuator and a capacitive sensor is considered in this work, which is a type of the precision nano-positioner. The maximum open-loop travel range and the open-loop resolution are respectively $[-10,120] \enspace {\mu}\text{m}$ and 0.2 nm. The E712 digital module performs as the amplifier to drive the actuator. The ranges of the amplifier's input and output are $[-3,13.5]$ V and $[-30,135]$ V respectively.

The unloaded plant is identified by the linear model below
\begin{equation}
    \begin{array}{ll}
        P(j\omega)=P_e(j\omega).P_m(j\omega),\\
        P_e(j\omega)=\frac{K_e}{\frac{j\omega}{\omega_e}+1}, P_m(j\omega)=\frac{K_m}{(\frac{j\omega}{\omega_m})^2+2\frac{\xi_m}{\omega_m}.j\omega+1}
    \end{array}
    \label{eq: Linear model}
\end{equation}
where $P$ is the frequency transfer function (TF) of the plant, and $P_e$ and $P_m$ are the TFs of the electrical and mechanical dynamics respectively. In (\ref{eq: Linear model}), $K_e,f_e$ denote the DC gain and the corner frequency of the electrical dynamics, and $K_m,\xi_m,\omega_m$ denote the DC gain, the damping ratio, and the undamped natural frequency of the mechanical dynamics respectively. The values of these parameters are provided in Table \ref{tb: plant parameters}. The Bode plot of the plant will be depicted later in Section \ref{sec: robust control design}. The mechanical dynamics appears as a high-peak resonance and the amplifier, once connected to the piezo-actuator, shows a first-order mode. $P$ is considered as the nominal model of the plant.
\begin{table}[hb]
    \begin{center}
        \caption{The values of the problem parameters}\label{tb: plant parameters}
        \begin{tabular}{ccc|ccc}
            Parameter & Value & Unit & Parameter & Value & Unit\\ \hline
            $K_e$ & 10 &  & $K_m$ & 0.4986 & $\frac{{\mu}\text{m}}{\text{V}}$\\
            $f_e$ & 935 & Hz & $f_m$ & 747 & Hz\\
            $\boldsymbol{\kappa}$ & 1.6165 & & $\xi_m$ & 0.0089 & \\ \hline
            $\phi_m$ & 60 & deg & $\phi_M$ & 71 & deg\\
        \end{tabular}
    \end{center}
\end{table}

In this work, we focus on the hysteresis nonlinearity in the plant. This nonlinearity is mostly due to the piezo material property that is frequency independent. Fig. \ref{fig: hysteresis curves, DC gain uncertainty} depicts the hysteresis characteristics based on the measurements from the actuator's input to the plant's output. The hysteresis curves for the different travel ranges are shown in Fig. \ref{fig: hysteresis curves, DC gain uncertainty} (a). A frequency-domain analysis indicates a 215\% of variation in the DC gain of the mechanical part as shown in Fig. \ref{fig: hysteresis curves, DC gain uncertainty} (b).

\begin{figure}
    \begin{center}
        \includegraphics[width=9cm,trim=0 181 0 134, clip]{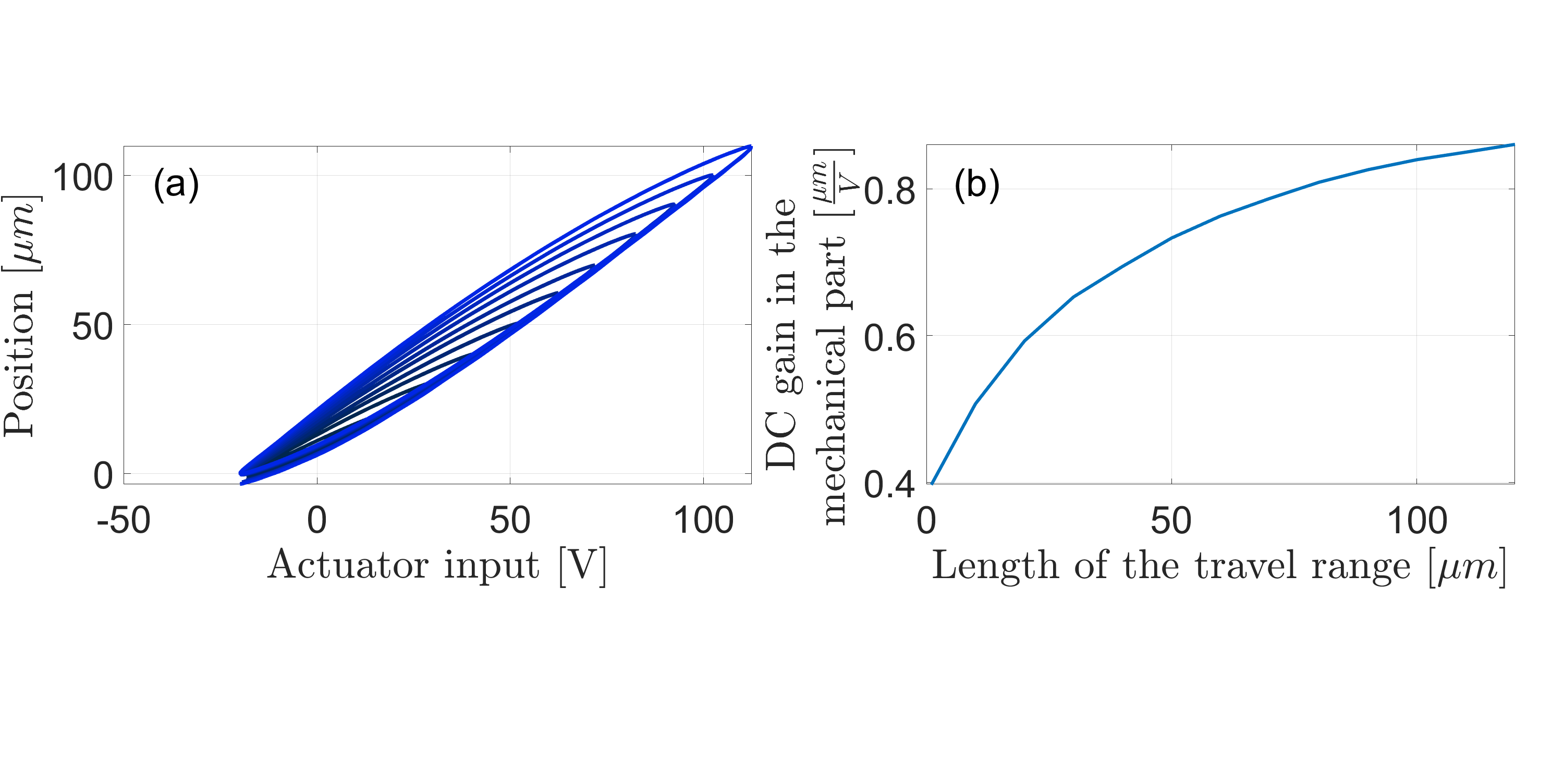}    
        \caption{Experimental study of the nonlinearity in the piezo setup; (a) Hysteresis curves; (b) The DC gain} 
    \label{fig: hysteresis curves, DC gain uncertainty}
    \end{center}
\end{figure}

An inverse model of the hysteresis is usually used to compensate for the hysteresis nonlinearity (\cite{NanopositioningTechs_Book_Piezo_Hysteresis}). However, due to the sensitivity of the compensation on the accuracy of the parameters of the inverse model, there is always a certain amount of remaining DC gain uncertainty. Therefore, the uncertain model below is considered for the plant where the uncertainty is representative of the remaining nonlinearity after any compensations:
\begin{equation}
    \begin{array}{ll}
        \mathcal{P}(j\omega,\kappa)=P_e(j\omega).\mathcal{P}_m(j\omega,\kappa),\\
        \mathcal{P}_m(j\omega,\kappa)=\kappa.P_m(j\omega), \kappa\in[1,\boldsymbol{\kappa}]\subset\mathbb{R}
    \end{array}
    \label{eq: Nonlinear model}
\end{equation}
In (\ref{eq: Nonlinear model}), $\mathcal{P}_m(\kappa)$ and $\mathcal{P}(\kappa)$ denote the dynamics of the mechanical part and the whole plant respectively where $\kappa$ is the uncertain DC gain. $\kappa=1,\boldsymbol{\kappa}$ respectively correspond to the nominal and worst cases where $\boldsymbol{\kappa}$ can be measured.
\subsection{Control objectives}
\label{subsec: control objectives}

Let consider the closed-loop TFs below with the open-loop TF \small $L(\kappa)=K_P.C.\mathcal{P}(\kappa), |L(j\omega_c(\kappa),\kappa)|=1$ \normalsize:
\begin{equation}
    \begin{array}{ll}
        S(j\omega,\kappa)=\frac{1}{1+L(j\omega,\kappa)}, T(j\omega,\kappa)=\frac{L(j\omega,\kappa)}{1+L(j\omega,\kappa)}
    \end{array}
    \label{eq: loop shaping, real error formulation}
\end{equation}
where the control loop is closed by the linear dynamical control $C$ and the proportional static control $K_P$ which are in series with the plant $\mathcal{P}(\kappa)$. The closed-loop TFs are respectively referred to by the sensitivity and the complementary sensitivity functions $S(\kappa),T(\kappa)$. $f_c$ denotes the cross-over frequency and determines the BW frequency. The argument $\kappa$ highlights the uncertainty in $f_c$ and the TFs. The BW average frequency is defined by:
\begin{equation}
    \begin{array}{ll}
        \Bar{f}_c=\frac{1}{f_c(\boldsymbol{\kappa})-f_c(1)}.\int_{\kappa=1}^{\kappa=\boldsymbol{\kappa}}f_c(\kappa)\,d\kappa
    \end{array}
    \label{eq: average bandwidth boundary}
\end{equation}
where $\Bar{f}_c$ is the average cross-over frequency. Next, the idea from the theory of the loop-shaping control is used to define the control objectives. For the robustness and precision purposes in the piezo-positioners, it is required to keep $|T(\kappa)|$ as close as possible to 0 dB for the frequencies till $f_c(\kappa)$. Thus, the open-loop gain $|L(\kappa)|$ needs to be respectively high and low before and after $f_c(\kappa)$ as much as possible, and, the open-loop phase ${\measuredangle}L(\kappa)$ needs to satisfy \small
\begin{equation}
    \begin{array}{ll}
        \phi(\kappa)\in[\phi_m,\phi_M]\subset\mathbb{R}_{++}, \phi(\kappa)={\measuredangle}L(j\omega_c(\kappa),\kappa)-(-180^{\circ})
    \end{array}
    \label{eq: constraint, phase margin bounds}
\end{equation} \normalsize
where $\phi$ denotes the phase margin and the lower and upper bounds $\phi_m$ and $\phi_M$ are given in Table. \ref{tb: plant parameters}.

A higher BW frequency, equivalently a higher $\bar{f}_c$, results in tracking references with higher speeds. Also, this usually increases the open-loop gain within the BW.
\section{Robust control based on integrated linear and complex-order elements}
\label{sec: control methodology}
In this section, a control methodology is established, which considers the control objectives that were explained in Section \ref{subsec: control objectives}. The formulations are provided for a general linear system with a DC gain uncertainty. The same notations are used as before. Robustness and maximum robust BW are defined below  as the first step.
\begin{figure}
    \begin{center}
        \includegraphics[width=1.6cm,angle=-90,trim=199 0 251 0, clip]{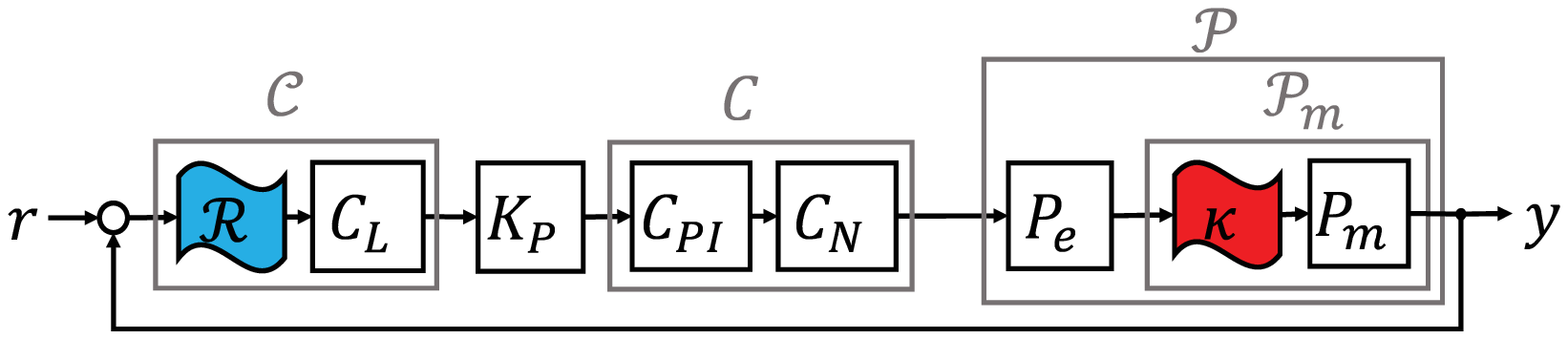}    
        \caption{The block-diagram of the control loop: the linear and nonlinear elements in the plant and the controller.} 
    \label{fig: archtecture}
    \end{center}
\end{figure}

\textbf{Definition 1:} The plant is given by the uncertain linear model $\mathcal{P}(j\omega,\kappa)=\kappa.P(j\omega)$ where $\kappa$ and $P(j\omega)$ are respectively the uncertain DC gain and the nominal linear model. The linear control $C(j\omega)$ and the proportional gain $K_P$ are in series with the plant within a feedback control loop such that the open-loop TF is formulated by $L(j\omega,\kappa)=K_P.C(j\omega).\mathcal{P}(j\omega,\kappa)$. For a set of finite bounds $\boldsymbol{\kappa},\phi_m,\phi_M$, the control $K_P.C(j\omega)$ is robust if phase margin constraint (\ref{eq: constraint, phase margin bounds}) is fulfilled for all $\kappa\in(1,\boldsymbol{\kappa})\subset\mathbb{R}$. Also, for the robust control $K_P.C(j\omega)$, the maximum robust BW is specified by the maximum average cross-over frequency $\bar{f}_c^*=\max\limits_{K_P\in\mathbb{R}}\{\bar{f}_c\}$ that is achieved by $K_P=K_P^*$.  $\square$

Next, an ideal control element is added to a given robust linear control to achieve a higher BW without jeopardizing the open-loop gain properties and the robustness. For this purpose, $L(\kappa)$ needs to fulfill certain conditions which are given in the following assumption.

\textbf{Assumption 1:} Let the linear plant $P(j\omega)$ and the linear control $\hat{C}(j\omega)$ be in series within a feedback control loop with the open-loop TF $\hat{L}(j\omega)=\hat{C}(j\omega).P(j\omega)$. The desired PM is given by (\ref{eq: constraint, phase margin bounds}) where the bounds $\phi_m,\phi_M$ are finite. There exist $\hat{C}(j\omega)=\hat{\textit{\textbf{C}}}(j\omega)$, $\hat{f}_a\in\mathbb{R}_{+}$, $\hat{f}_b,\hat{f}_B,\hat{f}_A\in\mathbb{R}_{++}$ such that \small
\begin{equation}
    \begin{array}{ll}
        {\measuredangle}\hat{L}\in[-180+\phi_m,-180+\phi_M)\subset\mathbb{R},\text{for}  f\in[\hat{f}_b,\hat{f}_B]\subset\mathbb{R}_{++},\\
        {\measuredangle}\hat{L}<-180+\phi_m\subset\mathbb{R},\enspace\text{for} \enspace f\in[\hat{f}_a,\hat{f}_b) \cup (\hat{f}_B,\hat{f}_A]\subset\mathbb{R}_{++},\\
        \frac{{\partial}|\hat{L}|}{{\partial}f}<0 \enspace\enspace \text{for} \enspace f\in\mathbb{R}_{++};\enspace\hat{f}_a<\hat{f}_b<\hat{f}_B<\hat{f}_A \quad\quad\quad \square
    \end{array}
    \label{eq: linear open-loop assumption}
\end{equation} \normalsize

For a given linear control which satisfies the conditions of Assumption 1, the DC gain of the control loop can be tuned by $K_P$ to achieve a robust linear control with the maximum robust BW if $\boldsymbol{\kappa}$ is less than a certain amount. The following proposition is given to obtain the maximum tolerable $\boldsymbol{\kappa}$ for the given linear control to be robust. $K_P$ is formulated to have the maximum robust BW.

\textbf{Proposition 1:} Let the given uncertain linear plant $\mathcal{P}(j\omega,\kappa)=\kappa.P(j\omega)$ and linear control $C(j\omega)$ be in series with the proportional gain $K_P\in\mathbb{R}_{++}$ within a control loop with the open-loop TF $L(j\omega,\kappa)=K_P.C(j\omega).\mathcal{P}(j\omega,\kappa)$.
Let the uncertain DC gain $\kappa$ of the plant be bounded as (\ref{eq: Nonlinear model}). Moreover, let Assumption 1 hold for the nominal linear plant $P(j\omega)$ and $\hat{C}(j\omega)=C(j\omega)$. $\boldsymbol{\hat{C}}(j\omega),\hat{f}_a,\hat{f}_b,\hat{f}_B,\hat{f}_A$ are respectively denoted by $\boldsymbol{C}(j\omega),f_a,f_b,f_B,f_A$. There exists $K_P=\textbf{\textit{K}}_{\textbf{\textit{P}}}$ such that the linear control $\textbf{\textit{K}}_{\textbf{\textit{P}}}.\textit{\textbf{C}}(j\omega)$ is robust under Definition 1 if and only if $\boldsymbol{\kappa}$ is bounded as:
\begin{equation}
    \begin{array}{ll}
        \boldsymbol{\kappa}{\leq}\frac{|\textbf{\textit{C}}(j\omega_b)|.|P(j\omega_b)|}{|\textbf{\textit{C}}(j\omega_B)|.|P(j\omega_B)|}
    \end{array}
    \label{eq: DC gain uncertainty bound}
\end{equation}
Moreover, for $\boldsymbol{\kappa}$ which satisfies (\ref{eq: DC gain uncertainty bound}) and the robust linear control $\textbf{\textit{K}}_{\textbf{\textit{P}}}.\textit{\textbf{C}}(j\omega)$, the maximum average cross-over frequency $\bar{\textbf{\textit{f}}}_\textbf{\textit{c}}^*$ under Definition 1 is achieved by setting $K_P=\textbf{\textit{K}}_{\textbf{\textit{P}}}^*$ where
\begin{equation}
    \begin{array}{ll}
        \textbf{\textit{K}}_{\textbf{\textit{P}}}^*=\frac{1}{\boldsymbol{\kappa}.|\textbf{\textit{C}}(j\omega_B)|.|P(j\omega_B)|}
    \end{array}
    \label{eq: proportional gain in linear control}
\end{equation}
Consequently, $\bar{\textbf{\textit{f}}}_\textbf{\textit{c}}^*$ is obtained by (\ref{eq: average bandwidth boundary}) with $f_c(\kappa)=\boldsymbol{f_c}^*(\kappa)$. $\boldsymbol{f_c}^*(\kappa)$ is the unique solution of the following equality where $\boldsymbol{L}^*(j\boldsymbol{\omega}(\kappa),\kappa)=\textbf{\textit{K}}_{\textbf{\textit{P}}}^*.\textbf{\textit{C}}(j\boldsymbol{\omega}).\mathcal{P}(j\boldsymbol{\omega}(\kappa))$:
\begin{equation}
    \begin{array}{ll}
        |\boldsymbol{L}^*(j\boldsymbol{\omega_c}^*(\kappa),\kappa)|=1
    \end{array}
    \label{eq: maximum nominal cross-over frequency}
\end{equation}
\\\textbf{Proof} (brief): Given the third condition of (\ref{eq: linear open-loop assumption}), formulation (\ref{eq: proportional gain in linear control}) sets $f_c(\boldsymbol{\kappa})=f_B$. $\boldsymbol{K_P}^*$ is the proportional gain candidate for the maximum robust BW according to the first and second conditions. Moreover, upper bound (\ref{eq: DC gain uncertainty bound}) implies that $\boldsymbol{K_P}^*$ is feasible regarding (\ref{eq: constraint, phase margin bounds}). $\square$

As implied by Proposition 1, there can be no feasible $K_P$ to have a robust linear control if $\boldsymbol{\kappa}$ is higher than a certain amount. Then, the linear control $C$ can be redesigned as one solution. Moreover, $\boldsymbol{{f}_c}^*(1)$, and accordingly, $\boldsymbol{\bar{f}_c}^*$ are shifted to lower values as $\boldsymbol{\kappa}$ increases. To increase the maximum robust BW, one idea is to push $f_B$, and accordingly $\boldsymbol{{f}_c}^*(\boldsymbol{\kappa})$, to a higher value. Thus, an extra phase is needed within a neighbor around $f=f_B$. In the context of linear control theory, an extra phase is only provided by lead elements. As a result, $|\boldsymbol{L}^*(\kappa)|$ increases after the BW or even decreases before the BW, which is in contrast with the desired properties of the open-loop gain. These limitations are due to the waterbed effect and the Bode gain-phase relationship in the linear control design.

A complex-order control is mathematically defined by: \small
\begin{equation}
    \begin{array}{ll}
        \tilde{\mathcal{C}}(j\omega)=e^{j(\frac{1-\tilde{\gamma}}{2})\arctan(\frac{\omega}{\tilde{\omega}_r})}, \enspace \tilde{\gamma}\in[-1,+1]\subset\mathbb{R}, \enspace \tilde{\omega}_r=2{\pi}\tilde{f}_r
    \end{array}
    \label{eq: complex-order control, frequency TF}
\end{equation} \normalsize
Fig. \ref{fig: archtecture} depicts the closed-loop control structure whose control elements will be defined in the next section. Once this control element is included into the control loop, a phase of $0^{\circ}$ to $90^{\circ}$, depending on the value of the parameter $\tilde{\gamma}$, is added to $\measuredangle{L(\kappa)}$ with the corner frequency $\tilde{f}_r$. It is done without any change in $|L(\kappa)|$. The following theorem states that it is always possible to achieve a higher robust BW without jeopardizing other control objectives by adding (\ref{eq: complex-order control, frequency TF}) to the given robust linear control.

\textbf{Theorem 1:} Let the given uncertain linear plant $\mathcal{P}(j\omega,\kappa)$ $=$$\kappa.P(j\omega)$ and linear control $C(j\omega)$ be in series with the proportional gain $\tilde{\mathcal{K}}_P\in\mathbb{R}_{++}$ and the complex-order control $\tilde{\mathcal{C}}(j\omega)$, defined by (\ref{eq: complex-order control, frequency TF}), within a control loop with the open-loop TF $\tilde{\mathcal{L}}(j\omega,\kappa)=\tilde{\mathcal{K}}_P.\tilde{\mathcal{C}}(j\omega).C(j\omega).\mathcal{P}(j\omega,\kappa)$.
Let the uncertain DC gain $\kappa$ of the plant be bounded as (\ref{eq: Nonlinear model}). Let Assumption 1 hold for the nominal linear plant $P(j\omega)$ and $\hat{C}(j\omega)=C(j\omega)$. $\boldsymbol{\hat{C}}(j\omega),\hat{f}_a,\hat{f}_b,\hat{f}_B,\hat{f}_A$ are respectively denoted by $\boldsymbol{C}(j\omega),f_a,f_b,f_B,f_A$. There exist $\tilde{\mathcal{f}}_b\in[0,f_b)\subset\mathbb{R}_{+},\tilde{\mathcal{f}}_B\in(f_B,\infty)\subset\mathbb{R}_{++},\tilde{\mathcal{K}}_P=\boldsymbol{\tilde{\mathcal{K}}_P}$ and 
$\tilde{\mathcal{C}}(j\omega)=\boldsymbol{\tilde{\mathcal{C}}}(j\omega)$ with $\tilde{\gamma}=\boldsymbol{\tilde{\gamma}},\tilde{f}_r=\boldsymbol{\tilde{f}_r}$ such that the control $\boldsymbol{\tilde{\mathcal{K}}_P}.\boldsymbol{\tilde{\mathcal{C}}}(j\omega).\boldsymbol{C}(j\omega)$ is robust under Definition 1 if and only if $\boldsymbol{\kappa}$ is bounded as:
\begin{equation}
    \begin{array}{ll}
        \boldsymbol{\kappa}{\leq}\frac{|\textbf{\textit{C}}(j\tilde{\mathcal{w}}_b)|.|P(j\tilde{\mathcal{w}}_b)|}{|\textbf{\textit{C}}(j\tilde{\mathcal{w}}_B)|.|P(j\tilde{\mathcal{w}}_B)|}
    \end{array}
    \label{eq: DC gain uncertainty bound; complex-order control}
\end{equation}
where the first condition of (\ref{eq: linear open-loop assumption}) holds for 
$P(j\omega),\hat{C}(j\omega)=\tilde{\mathcal{C}}(j\omega).C(j\omega)$ with $\boldsymbol{\hat{C}}(j\omega),\hat{f}_b,\hat{f}_B$ respectively replaced by $\boldsymbol{\tilde{\mathcal{C}}}(j\omega).\boldsymbol{C}(j\omega),\tilde{\mathcal{f}}_b,\tilde{\mathcal{f}}_B$. Moreover, for $\boldsymbol{\kappa}$ which satisfies (\ref{eq: DC gain uncertainty bound; complex-order control}) and the roust control $\boldsymbol{\tilde{\mathcal{K}}_P}.\boldsymbol{\tilde{\mathcal{C}}}(j\omega).\boldsymbol{C}(j\omega)$, the maximum average cross-over frequency $\boldsymbol{\bar{\tilde{\mathcal{f}}}_c}^*$ under Definition 1 is achieved by setting $\tilde{\mathcal{K}}_P=\tilde{\boldsymbol{\mathcal{K}}}_P^*$ where
\begin{equation}
    \begin{array}{ll}
        \boldsymbol{\tilde{\mathcal{K}}}_P^*=\frac{1}{\boldsymbol{\kappa}.|\textbf{\textit{C}}(j\tilde{\mathcal{w}}_B)|.|P(j\tilde{\mathcal{w}}_B)|}
    \end{array}
    \label{eq: proportional gain in complex-order control}
\end{equation}
Consequently, $\boldsymbol{\bar{\tilde{\mathcal{f}}}_c}^*$ is obtained by (\ref{eq: average bandwidth boundary}) with $f_c(\kappa)=\boldsymbol{\tilde{\mathcal{f}}_c}^*(\kappa)$. $\boldsymbol{\tilde{\mathcal{f}}_c}^*(\kappa)$ is the unique solution of the following equality where $\boldsymbol{\tilde{\mathcal{L}}}^*(j\boldsymbol{\tilde{\mathcal{w}}}(\kappa),\kappa)=\boldsymbol{\tilde{\mathcal{K}}_P}^*.\boldsymbol{\tilde{\mathcal{C}}}(j\boldsymbol{\tilde{\mathcal{w}}}).\textbf{\textit{C}}(j\boldsymbol{\tilde{\mathcal{w}}}).\mathcal{P}(j\boldsymbol{\tilde{\mathcal{w}}}(\kappa))$:
\begin{equation}
    \begin{array}{ll}
        |\boldsymbol{\tilde{\mathcal{L}}}^*(j\boldsymbol{\tilde{\mathcal{w}}_c}^*(\kappa),\kappa)|=1
    \end{array}
    \label{eq: maximum cross-over frequency; complex-order control}
\end{equation}
Moreover, the following inequalities consequently hold:
\begin{equation}
    \begin{array}{ll}
        \boldsymbol{\tilde{\mathcal{f}}_c}^*(\kappa)>\boldsymbol{f_c}^*(\kappa), \enspace \boldsymbol{\bar{\tilde{\mathcal{f}}}_c}^*>\boldsymbol{\bar{f}_c}^*
    \end{array}
    \label{eq: maximum (average) cross-over frequency improved by complex-order control}
\end{equation}
$\boldsymbol{f_c}^*(\kappa)$ and $\boldsymbol{\bar{f}_c}^*$ are associated with $\boldsymbol{K_P}^*$ for the linear control $\boldsymbol{K_P}^*.\boldsymbol{C}(j\omega)$ according to Proposition 1.\\
\textbf{Proof} (brief): Formulations (\ref{eq: proportional gain in complex-order control}), (\ref{eq: maximum cross-over frequency; complex-order control}) are derived similar to the proof of Proposition 1. Because of the added phase by the complex-order element (\ref{eq: complex-order control, frequency TF}) and with regard to the conditions in (\ref{eq: linear open-loop assumption}), a parametrization of (\ref{eq: complex-order control, frequency TF}) is possible to have $\boldsymbol{\tilde{\mathcal{f}}_B}>\boldsymbol{f_B},\boldsymbol{\tilde{\mathcal{f}}_b}<\boldsymbol{f_b}$. Therefore, (\ref{eq: maximum (average) cross-over frequency improved by complex-order control}) is concluded. $\square$

Inequalities (\ref{eq: maximum (average) cross-over frequency improved by complex-order control}) imply that, mathematically, the maximum robust BW can always increase after designing a robust linear control for a linear plant with a bounded DC gain uncertainty. This is done by adding ideal complex-order element (\ref{eq: complex-order control, frequency TF}) to the controller.

\textbf{Remark 1:} The upper bound in (\ref{eq: DC gain uncertainty bound; complex-order control}) is less limiting than the upper bound in (\ref{eq: DC gain uncertainty bound}) because of the third condition in (\ref{eq: linear open-loop assumption}). It means that robustness against higher DC gain uncertainty is possible by the proposed control scheme compared to the given linear control.
\section{Robust motion control design for piezo-positioner with hysteresis nonlinearity}\label{sec: robust control design}
The aim is to improve robust motion control deign to achieve higher precision and BW in performance of piezo-actuated nano-positioning systems. The model of the piezo-positioner is considered as in Section \ref{subsec: system description} and the desired performance is taken into account by the control objectives which were explained in Section \ref{subsec: control objectives}. The control methodology which was proposed in Section \ref{sec: control methodology} is realized to achieve the improvements.

In this section, a robust linear motion control is designed based on Proposition 1 as the first step. A set of simple linear elements is selected to focus on the main contributions. Then, a robust nonlinear motion control scheme is proposed to utilize Theorem 1.
\subsection{Robust linear motion control design}
\label{subsec: linear control elements}
The high-peak resonance in the mechanical dynamics is limiting to achieve a high BW. The notch filter \small $C_N(j\omega)=\frac{(\frac{j\omega}{\omega_{m}})^2+2\frac{\xi_{m}}{\omega_{m}}.j\omega+1}{(\frac{j\omega}{\omega_{m}})^2+2\frac{\xi_{N}}{\omega_{m}}.j\omega+1}$ \normalsize is used to replace the resonance with an over-damped second-order dynamics which is determined by the damp ratio $\xi_N$ and the frequency $f_m$.

According to the loop-shaping rule, at least one integrator is necessary to provide a high open-loop gain before $f_c(\kappa)$. Moreover, the open-loop phase needs to fulfill (\ref{eq: constraint, phase margin bounds}) within a range of frequency. Therefore, the second-order PI control $C_{\textit{PI}}(j\omega)=(1+\frac{\omega_{i}}{j\omega}).(1+\frac{\omega_{i}'}{j\omega})$ is included where the proportional gain is equal to 1 and the integrator coefficients $f_{i},f_{i}'$ are the corner frequencies. 

\begin{figure}
    \begin{center}
        \includegraphics[width=9cm,trim=0 145 0 170, clip]{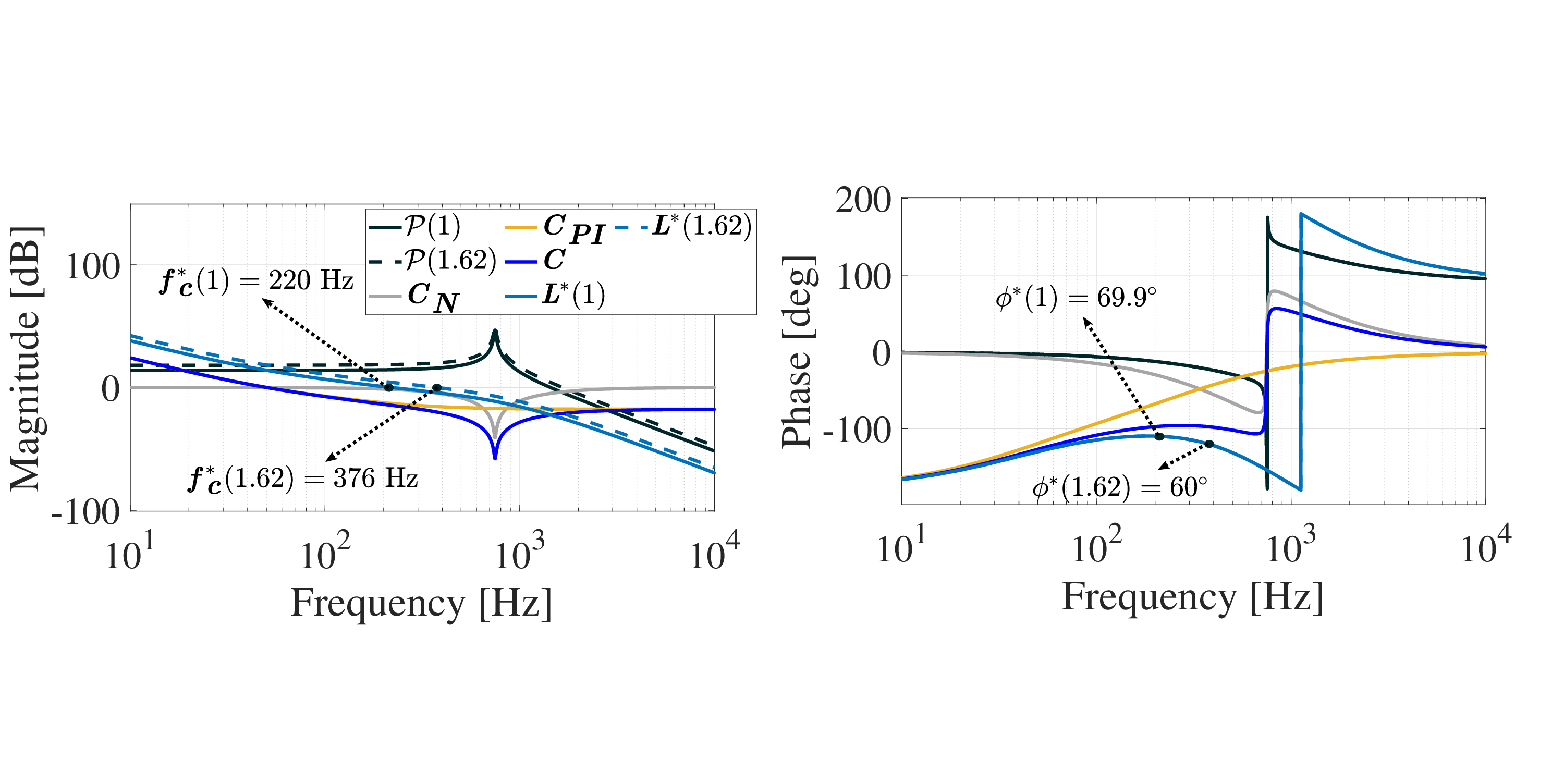}    
        \caption{The Bode plots of the plant and the open loop in the nominal and worst cases together with the Bode plots of the robust linear control and its elements.} 
    \label{fig: robust linear control design}
    \end{center}
\end{figure}

Accordingly, the linear control $C$ and the corresponding open-loop TF $L(\kappa)$ are formulated by \small
\begin{equation}
    \begin{array}{ll}
        L(j\omega,\kappa)=K_{P}.C(j\omega).\mathcal{P}_(j\omega,\kappa), C(j\omega)=C_{\textit{PI}}(j\omega).C_N(j\omega)
    \end{array}
    \label{eq: linear control, linear open-loop}
\end{equation} \normalsize
The Bode plots of the linear control $C$ and $L(\kappa)$ are depicted in Fig. \ref{fig: robust linear control design}. Based on Proposition 1, the proportional control $K_{P}$ is tuned to achieve the maximum robust BW for the given control structure in (\ref{eq: linear control, linear open-loop}) and the given uncertainty bound. The result of this tuning $\boldsymbol{K_P}^*$ as well as the other control parameters are given in Table. \ref{tb: control parameters}. As indicated in Fig. \ref{fig: robust linear control design}, the maximum nominal and worst-case cross-over frequencies $\boldsymbol{f_c}^*(1)=220\enspace\text{Hz},\boldsymbol{f_c}^*(\boldsymbol{\kappa})=376\enspace\text{Hz}$ are achieved that satisfy PM constraints (\ref{eq: constraint, phase margin bounds}).
\subsection{Robust nonlinear motion control design based on CgLp reset control}
\label{subsec: CgLp design}
It was shown in Section \ref{subsec: linear control elements} that the uncertainty in the DC gain is limiting to achieve a higher cross-over frequency. Around the frequency $\boldsymbol{f_c}^*(\boldsymbol{\kappa})$, ${\measuredangle}L$ drops with a high negative slope, which supports this limitation. Based on the iso-damping control, as a solution in the literature, a linear lead element adds phase to modify the negative slope. On the other hand, the lead element increases the slope of $|L(\kappa)|$. This is in contrast with the desired property for $|L(\kappa)|$ as one of the control objectives.

The idea is to increase the slope of the open-loop phase after $\boldsymbol{f_c}^*(\kappa)$ to achieve the robustness at higher frequencies without a negative effect on the open-loop gain. The designed linear control in Section (\ref{subsec: linear control elements}) and DC gain uncertainty satisfy Assumption 1 and (\ref{eq: DC gain uncertainty bound}) respectively. Thus, the conditions of Theorem 1 hold and there exists a feasible complex-order control with definition (\ref{eq: complex-order control, frequency TF}) to increase the BW while the closed-loop system is robust. It is not possible to realize complex-order control (\ref{eq: complex-order control, frequency TF}) by causal linear elements.

CgLp reset control is a nonlinear control that approximately realizes (\ref{eq: complex-order control, frequency TF}) (\cite{valerio2019ComplexOrderCgLp}). Fig. \ref{fig: archtecture} indicates the considered CgLp control architecture. It consists of a first-order reset element (FORE) followed by a first-order lead filter. The sequence is selected such that noise, which will exist in practice, is not amplified in the input of the FORE. The FORE is recognized by the base linear dynamics $\mathcal{R}_{\textit{bl}}=\frac{1}{j\frac{\omega}{\omega_r}+1}$ and a zero-crossing resetting action.

To study the dynamics of the FORE in the frequency domain, the SIDF and HOSIDF tools are used which are respectively associated with the first harmonic and the higher-order harmonics of the output in response to sinusoidal input. The SIDF of the CgLp control is given by (\cite{saikumar2019CgLp}): \small
\begin{equation}
    \begin{array}{ll}
        \mathcal{C}_1(j\omega)=\mathcal{R}_1(j\omega).C_{L}(j\omega), \enspace C_{L}(j\omega)=\frac{j\frac{\omega}{\alpha\omega_r}+1}{j\frac{\omega}{\omega_f}+1},\\
        \mathcal{R}_1(j\omega)=\frac{1}{j\frac{\omega}{\omega_r}+1}.(1+j\frac{2\omega^2(1-\gamma)(1+e^{-\frac{\pi}{\omega}\omega_r})}{\pi(1+{\gamma}e^{-\pi\frac{\omega_r}{\omega}})(\omega^2+\omega_r^2)})
    \end{array}
    \label{eq: CgLp SIDF}
\end{equation} \normalsize
where $\mathcal{R}_1$ and $C_{L}$ are respectively the SIDF and the TF of the FORE and the lead filter. $f_r,f_f$ are the corner frequencies and $\gamma\in[-1,+1]\subset\mathbb{R},\alpha\in[1,\infty)\subset\mathbb{R}_{++}$ are the resetting and corresponding correction factors.

Based on SIDF (\ref{eq: CgLp SIDF}), the CgLp element adds phase lead to the open-loop TF up to $90^{\circ}$ within the frequency range which is determined by $f_r,f_f$. There is almost no effect on the open-loop gain. Therefore, (\ref{eq: CgLp SIDF}) is a realization of (\ref{eq: complex-order control, frequency TF}).

Since the CgLp control is a nonlinear element, it is important to be aware of the control nonlinearity effects. This nonlinearity appears as higher-order harmonics. Under certain conditions (\cite{saikumar2021loopshaping}), the sensitivity and complementary sensitivity functions together with the open-loop TFs are given as (also see \cite{hou2020tuning_CgLp_HOSIDF}): \small
\begin{subequations}
\begin{equation}
    \begin{array}{ll}
        \mathcal{S}_1(j\omega,\kappa)=\frac{1}{1+\mathcal{L}_1(j\omega,\kappa)}, \mathcal{T}_1(j\omega,\kappa)=\frac{\mathcal{L}_1(j\omega,\kappa)}{1+\mathcal{L}_1(j\omega,\kappa)},\\
        \mathcal{L}_1(j\omega,\kappa)=\mathcal{C}_1(j\omega).\mathcal{K}_P.C(j\omega).\mathcal{P}(j\omega,\kappa)
    \end{array}
    \label{eq: loop shaping based on DF, real error formulation}
\end{equation}
\begin{equation}
    \begin{array}{ll}
        \mathcal{S}_n(j\omega,\kappa)=-\mathcal{T}_n(j\omega,\kappa)=-S_{bl}(jn\omega,\kappa).\mathcal{L}_n(j\omega,\kappa).\\
        \frac{(\mathcal{S}_1(j\omega,\kappa))^n}{|\mathcal{S}_1(j\omega,\kappa)|^{n-1}}.\frac{(R(j\omega))^{n-1}}{|R(j\omega)|^{n-1}},S_{\textit{bl}}(j\omega,\kappa)=\frac{1}{1+\mathcal{L}_{\textit{bl}}(j\omega,\kappa)},\\
        \mathcal{L}_{\textit{bl}}(j\omega)=\mathcal{C}_{\textit{bl}}(j\omega).\mathcal{K}_P.C(j\omega).\mathcal{P}(j\omega,\kappa), \mathcal{C}_{\textit{bl}}(j\omega)=\\
        \mathcal{R}_{\textit{bl}}(j\omega).C_L(j\omega),\mathcal{L}_n(j\omega,\kappa)=\mathcal{C}_n(j\omega).\mathcal{K}_P.C(jn\omega).\\
        \mathcal{P}(jn\omega,\kappa),\mathcal{C}_n(j\omega)=\mathcal{R}_n(j\omega).C_L(jn\omega), \mathcal{R}_n(j\omega)=\\
        \begin{cases}
        \frac{1}{j\frac{n\omega}{\omega_r}+1}.(j\frac{2\omega^2(1-\gamma)(1+e^{-\frac{\pi}{\omega}\omega_r})}{\pi(1+{\gamma}e^{-\pi\frac{\omega_r}{\omega}})(\omega^2+\omega_r^2)}) & n: \text{odd}\\
        0 & n: \text{even}
        \end{cases}
    \end{array}
    \label{eq: Higher-order sensitivity functions}
\end{equation}
\begin{equation}
    \begin{array}{ll}
        \mathcal{S}_{\infty}(j\omega)=\max{\{\Sigma_{i=1}^{\infty}|\mathcal{S}_i(j\omega,\kappa)|\sin{(n{\omega}t+{\measuredangle}\mathcal{S}_i(j\omega,\kappa))}\}},\\
        \mathcal{T}_{\infty}(j\omega)=\max{\{\Sigma_{i=1}^{\infty}|\mathcal{T}_i(j\omega,\kappa)|\sin{(n{\omega}t+{\measuredangle}\mathcal{T}_i(j\omega,\kappa))}\}}
    \end{array}
    \label{eq: Pseudo sensitivity functions}
\end{equation}
\end{subequations} \normalsize
where $\mathcal{S}_1(\kappa),\mathcal{T}_1(\kappa)$ and $\mathcal{S}_n(\kappa),\mathcal{T}_n(\kappa)$ are the sensitivity and complementary sensitivity functions based on SIDF and HOSIDF respectively. $\mathcal{S}_{\infty}(\kappa),\mathcal{T}_{\infty}$ are the pseudo sensitivity and pseudo complementary sensitivity functions that approximate the real sensitivity and complementary sensitivity functions. To have a valid SIDF-based CgLp reset control design, the effects of the higher-order harmonics need to be sufficiently low. Therefore, $|\mathcal{S}_n(\kappa)|,|\mathcal{T}_n(\kappa)|$ must be sufficiently lower than $|\mathcal{S}_1(\kappa)|,|\mathcal{T}_1(\kappa)|$. As a result, $|\mathcal{S}_{\infty}(\kappa)|,|\mathcal{T}_{\infty}(\kappa)|$ and $|\mathcal{S}_1(\kappa)|,|\mathcal{T}_1(\kappa)|$ must be almost the same. In (17), $\mathcal{R},\mathcal{C},\mathcal{L}$ respectively denote the frequency responses of the FORE, CgLp control, and open loop where subscripts $1,n,\textit{bl}$ respectively refer to SIDF-based, HOSIDF-based, and base linear analyses.

A constrained optimization problem is presented to optimize the parameters of the CgLp control regarding the control objectives in Section \ref{subsec: control objectives}. The cross-over frequency $\mathcal{f}_c$ is the profit function. (\ref{eq: constraint, phase margin bounds}) is the constraint for all $\kappa$ with the bounds as in (\ref{eq: Nonlinear model}). Thus, the control parameters $f_r=\boldsymbol{f_r}^*,f_f=\boldsymbol{f_f}^*,\gamma=\boldsymbol{\gamma}^*$ are selected where $\boldsymbol{f_r}^*,\boldsymbol{f_f}^*,\boldsymbol{\gamma}^*$ are the solution of the following problem with $\mathcal{w}_c=\{\mathcal{w}_c(1),\beta.\mathcal{w}_c(1))\}$:
\begin{equation}
    \begin{aligned}
        \max_{\gamma,f_r,f_f} \quad & \mathcal{f}_c(1) \\
        \textrm{s.t.} \quad & \measuredangle{\mathcal{L}_1}(j\mathcal{w}_c)\in[-180+\phi_m,-180+\phi_M]\\
        & \gamma\in[\gamma_m,1]\subset\mathbb{R}, \enspace f_r>\nu.f_f, \enspace f_f<\mathcal{f}_M
    \end{aligned}
    \label{eq: Constrained Optimization Problem}
\end{equation}
The phase constraint takes care of (\ref{eq: constraint, phase margin bounds}). $\beta\in\mathbb{R}$ is roughly determined such that $\mathcal{f}_c(\boldsymbol{\kappa}){\approx}\beta.\mathcal{f}_c(1)$ by looking at $\boldsymbol{L}^*$. The tuning parameters $\gamma_m\in[-1,1]\subset\mathbb{R},\nu\in(0,1)\subset\mathbb{R}_{++}$ respectively limit the intensity and the range of the introduced nonlinearity by the FORE. $\mathcal{f}_M\in\mathbb{R_{++}}$ is set regarding the available sampling rate in practice. The optimal proportional gain $\boldsymbol{\mathcal{K}_P}^*$ is then obtained by \small
\begin{equation}
    \begin{array}{ll}
        \boldsymbol{\mathcal{K}_P}^*=\frac{1}{|\boldsymbol{\mathcal{C}(j\mathcal{w}_c}^*(\boldsymbol{\kappa}))|.|\boldsymbol{C(j\mathcal{w}_c}^*(\boldsymbol{\kappa}))|.|\mathcal{P}(\boldsymbol{j\mathcal{w}_c}^*(\boldsymbol{\kappa}),\boldsymbol{\kappa})|}
    \end{array}
    \label{eq: proportional gain in nonlinear control}
\end{equation} \normalsize
A manual retuning of $\boldsymbol{\mathcal{K}_P}^*$ can be done finally to maximize $\mathcal{f}_c^*(\boldsymbol{\kappa})$ accurately regarding the $\beta$ approximation.

\begin{table}[hb]
    \begin{center}
        \caption{The values of the control parameters}\label{tb: control parameters}
        \begin{tabular}{ccc|ccc}
            Parameter & Value & Unit & Parameter & Value & Unit \\ \hline
            $f_i$ & 300 & Hz & $f_i'$ & 40 & Hz \\
            $\xi_N$ & 1 & & $\boldsymbol{K_P}^*$ & 0.1303 & $\frac{V}{{\mu}m}$ \\
            $\boldsymbol{f_c}^*(1)$ & 220 & Hz & $\boldsymbol{f_c}^*(\boldsymbol{\kappa})$ & 376 & Hz \\
            $\boldsymbol{\phi_{\textit{\textbf{c}}}}^*(1)$ & 70 & deg & $\boldsymbol{\phi_{\textit{\textbf{c}}}}^*(\boldsymbol{\kappa})$ & 60 & deg \\ \hline
            $\boldsymbol{f_r}^*$ & 324 & Hz & $\boldsymbol{f_f}^*$ & 4.206 & KHz \\
            $\gamma,\alpha$ & 0.3,1.21 &  & $\boldsymbol{\mathcal{K}_P}^*$ & 0.1645 & $\frac{V}{{\mu}m}$ \\
            $\boldsymbol{\mathcal{f}_c}^*(1)$ & 269 & Hz & $\boldsymbol{\mathcal{f}_c}^*(\boldsymbol{\kappa})$ & 441 & Hz \\
            $\boldsymbol{\varphi_c}^*(1)$ & 69 & deg & $\boldsymbol{\varphi_c}^*(\boldsymbol{\kappa})$ & 60 & deg
        \end{tabular}
    \end{center}
\end{table}

\section{Simulation results}\label{sec: simulation results}
The nonlinear motion control scheme, presented in Section \ref{subsec: CgLp design}, is designed for the considered piezo-positioner. The control parameters and outcome are available in Table. \ref{tb: control parameters} in case of $\beta=\frac{\boldsymbol{f_c}^*(\boldsymbol{\kappa})}{\boldsymbol{f_c}^*(1)},\gamma_m=0.3,\nu=\frac{1}{13},f_M=7\text{KHz}$. According to Fig. \ref{fig: Bode, Sensitivity}(a), the open-loop phase is brought up after $\boldsymbol{f_c}^*(\boldsymbol{\kappa})$. Consequently, $\boldsymbol{\mathcal{f}_c}^*(1),\boldsymbol{\mathcal{f}_c}^*(\boldsymbol{\kappa})$ are respectively 22.3\% and 17.3\% higher than $\boldsymbol{f_c}^*(1),\boldsymbol{f_c}^*(\boldsymbol{\kappa})$, meaning that an increase of about 20\% has been achieved in the maximum robust BW. Fig. \ref{fig: Bode, Sensitivity}(b) shows the achieved sensitivity function in the robust nonlinear motion control outcome is lower than the case of the robust linear motion control till around $f_m$ for all $\kappa{\in}[1,\boldsymbol{\kappa}]$. Fig. \ref{fig: Bode, Sensitivity}(c) shows the $\pm1$ dB criterion on the closed-loop gain is almost fulfilled by the nonlinear control. The outcome is also compared with a non-robust linear control. In this case, $f_c(1)=\boldsymbol{\mathcal{f}_c}^*(1)$ while the $\pm1$ dB criterion is violated.
\begin{figure}
    \begin{center}
        \includegraphics[width=9cm,trim=0 4 0 1, clip]{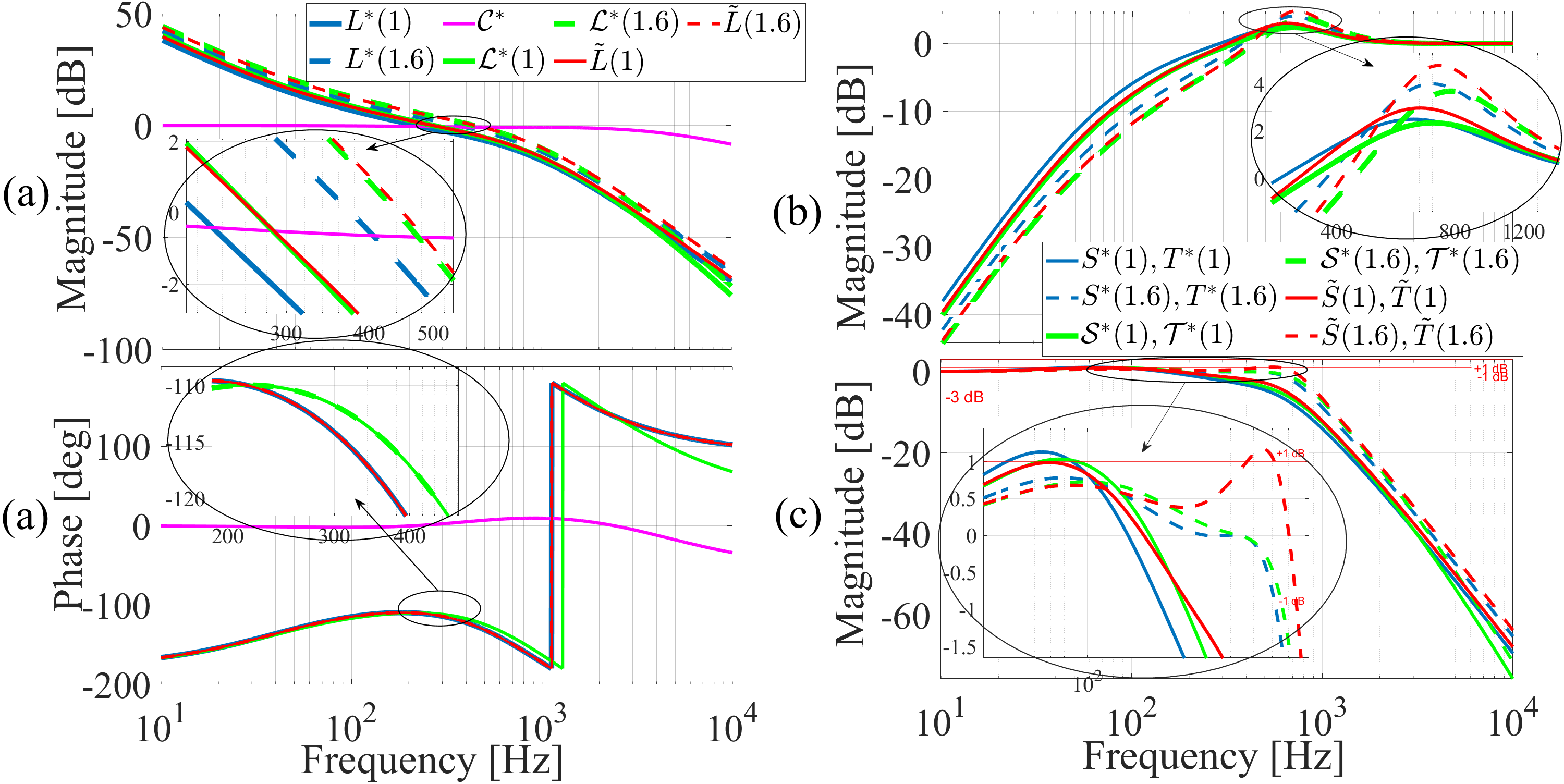}    
        \caption{SIDF-based comparison of robust nonlinear control with linear controllers; (a) Open loop, (b) Sensitivity function, (c) Complementary sensitivity function.} 
    \label{fig: Bode, Sensitivity}
    \end{center}
\end{figure}
\begin{figure}
    \begin{center}
        \includegraphics[width=9cm,trim=0 3 0 5, clip]{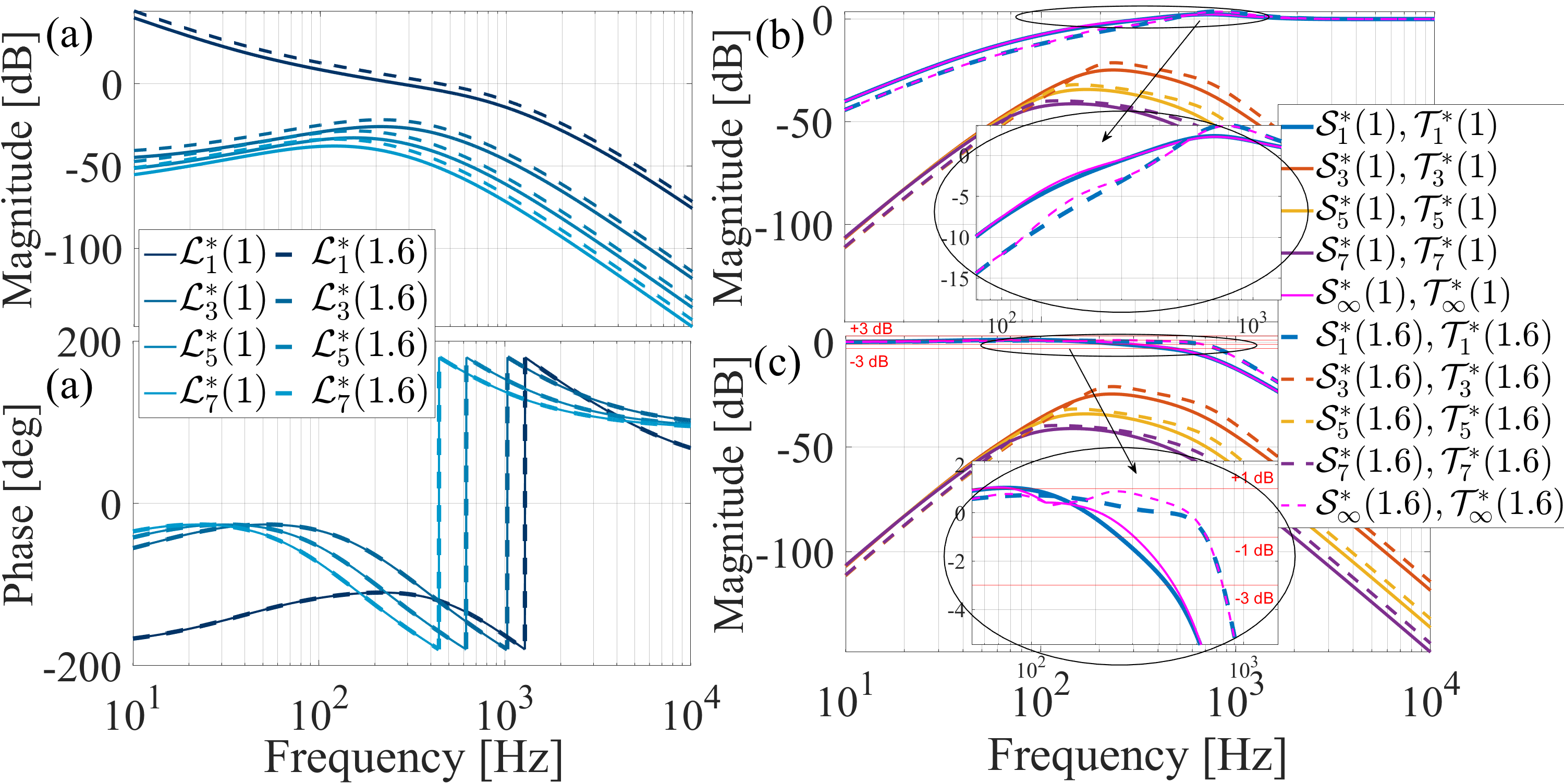}    
        \caption{The HOSIDF-based validation of the designed control; (a) Open loop, (b) Sensitivity function, (c) Complementary sensitivity function.} 
    \label{fig: HOSIDF; CgLp, L, Sensitivity, Complementary, Pseudo}
    \end{center}
\end{figure}
\begin{figure}
    \begin{center}
        \includegraphics[width=8.8cm,trim=0 8 0 3, clip]{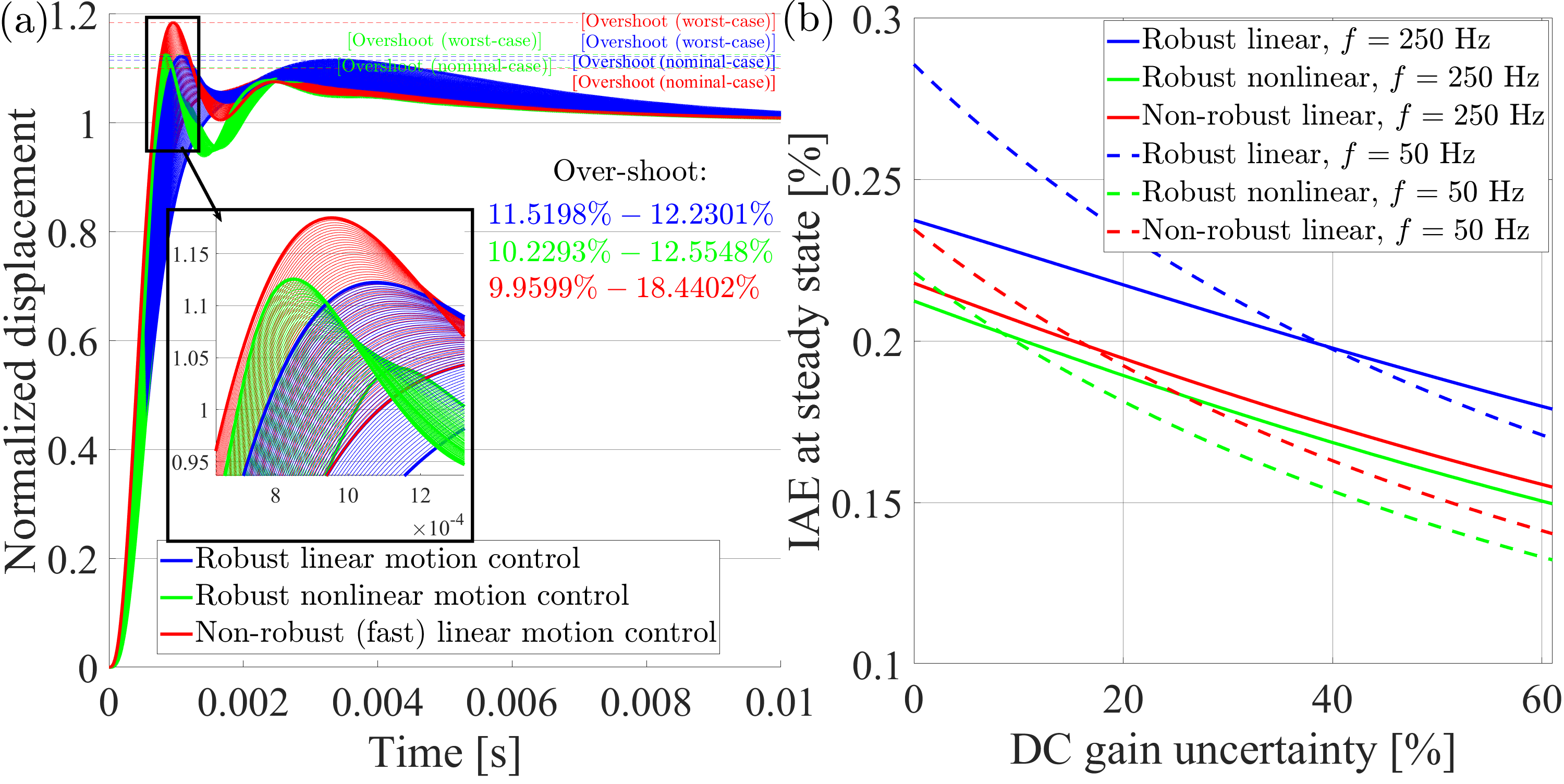}    
        \caption{(a) Step response comparison; (b) Sinusoid response comparison}
    \label{fig: Step response comparison}
    \end{center}
\end{figure}

Fig. \ref{fig: HOSIDF; CgLp, L, Sensitivity, Complementary, Pseudo} shows the open-loop HOSIDFs magnitudes are sufficiently low such that $|\mathcal{S}_{\infty}(\kappa)|{\approx}|\mathcal{S}_1(\kappa)|$, $|\mathcal{T}_{\infty}(\kappa)|{\approx}|\mathcal{T}_1(\kappa)|$. Thus, the SIDF-based analysis is validated in the frequency domain.

Fig. \ref{fig: Step response comparison} highlights the improvement compared to the linear controllers. The responses are simulated over the full range of the DC gain uncertainty as a representative of different travel ranges. Fig. \ref{fig: Step response comparison}(a) shows the achieved rise-time is less than the robust linear control outcome, indicating the improvement in the BW. The over-shoot is kept within 10\% to 13\% while it varies from 10\% to 18\% in the non-robust linear control outcome, indicating robustness is conserved although the maximum robust BW has been increased, which is not possible by linear control. Fig. \ref{fig: Step response comparison}(b) highlights the improvement in the precision by the integral of absolute error (IAE) index. In practice, additional filters might be needed regarding the actuator's minimum rise time and the amplifier's slew rate and sampling rate.

\section{Conclusion}
In this work, a robust nonlinear control scheme based on a CgLp reset control was presented for the motion control of a piezo-stage. The robustness is against the DC gain uncertainty in the plant. The uncertainty is representative of the hysteresis nonlinearity in the piezo material. It was shown that the cross-over frequency is shifted to a higher value while satisfying the PM constraints by adding the CgLp element to a given robust linear control. This is done without sacrificing the desired properties of the open-loop gain. The results for the case of 62\% uncertainty, which is an exaggerated nonlinearity effect, indicated about 20\% increase in the average BW while keeping the closed-loop gain within the $\pm1$ dB bound around 0 dB, meaning that the maximum speed of variation in the reference can be averagely 20\% faster in the case of the designed robust nonlinear control compared to the given robust linear control while the stage can still track the reference signal with the $\pm1$ dB criterion. A significant improvement in precision was shown. Thus, a high-performant tracking is achieved for the piezo-actuated nano-positioner in case of references with different amplitudes.

The relevant and potential future research are: Investigating the time response of the control effort and designing a new architecture by adding extra filters if needed; Experimental validations; Incorporating feedforward control; Looking into the disturbance rejection problem.

\begin{ack}
The authors express their sincere gratitude to both Mathias Winter, Head of Piezo System \& Drive Technology, and Dr.-Ing. Simon Kapelke, Head of Piezo Fundamental Technology, from Physik Instrumente (PI) SE \& Co. KG, for their invaluable collaboration in providing technical insights concerning the system and its applications.
\end{ack}

\bibliography{ifacconf}             

\begin{thebibliography}{9}
\providecommand{\natexlab}[1]{#1}
\providecommand{\url}[1]{\texttt{#1}}
\providecommand{\urlprefix}{URL }
\expandafter\ifx\csname urlstyle\endcsname\relax
  \providecommand{\doi}[1]{doi:\discretionary{}{}{}#1}\else
  \providecommand{\doi}{doi:\discretionary{}{}{}\begingroup \urlstyle{rm}\Url}\fi

\bibitem[{Clegg(1958)}]{clegg1958nonlinear}
Clegg, J.C. (1958).
\newblock A nonlinear integrator for servomechanisms.
\newblock \emph{Transactions of the American Institute of Electrical Engineers, Part II: Applications and Industry}, 77(1), 41--42.

\bibitem[{Flores et~al.(2020)Flores, Mu{\~n}oz, Monje, Milan{\'e}s, and Lu}]{IsoDamping_Car_Bode}
Flores, C., Mu{\~n}oz, J., Monje, C.A., Milan{\'e}s, V., and Lu, X.Y. (2020).
\newblock Iso-damping fractional-order control for robust automated car-following.
\newblock \emph{Journal of advanced research}, 25, 181--189.

\bibitem[{Hou et~al.(2020)Hou, Dastjerdi, Saikumar, and Hosseinnia}]{hou2020tuning_CgLp_HOSIDF}
Hou, X., Dastjerdi, A.A., Saikumar, N., and Hosseinnia, S.H. (2020).
\newblock Tuning of ‘constant in gain lead in phase (cglp)’reset controller using higher-order sinusoidal input describing function (hosidf).
\newblock In \emph{2020 Australian and New Zealand Control Conference (ANZCC)}, 91--96. IEEE.

\bibitem[{Ru et~al.(2016)Ru, Liu, Sun et~al.}]{NanopositioningTechs_Book_Piezo_Hysteresis}
Ru, C., Liu, X., Sun, Y., et~al. (2016).
\newblock Nanopositioning technologies.
\newblock \emph{Fundamentals and applications}.

\bibitem[{Saikumar et~al.(2021)Saikumar, Heinen, and HosseinNia}]{saikumar2021loopshaping}
Saikumar, N., Heinen, K., and HosseinNia, S.H. (2021).
\newblock Loop-shaping for reset control systems: A higher-order sinusoidal-input describing functions approach.
\newblock \emph{Control Engineering Practice}, 111, 104808.

\bibitem[{Saikumar et~al.(2019)Saikumar, Sinha, and HosseinNia}]{saikumar2019CgLp}
Saikumar, N., Sinha, R.K., and HosseinNia, S.H. (2019).
\newblock “constant in gain lead in phase” element--application in precision motion control.
\newblock \emph{IEEE/ASME Transactions on Mechatronics}, 24(3), 1176--1185.

\bibitem[{Val{\'e}rio et~al.(2019)Val{\'e}rio, Saikumar, Dastjerdi, Karbasizadeh, and HosseinNia}]{valerio2019ComplexOrderCgLp}
Val{\'e}rio, D., Saikumar, N., Dastjerdi, A.A., Karbasizadeh, N., and HosseinNia, S.H. (2019).
\newblock Reset control approximates complex order transfer functions.
\newblock \emph{Nonlinear Dynamics}, 97, 2323--2337.

\bibitem[{Yonezawa et~al.(2024)Yonezawa, Yonezawa, Yahagi, and Kajiwara}]{IsoDamping_DataDriven_New}
Yonezawa, A., Yonezawa, H., Yahagi, S., and Kajiwara, I. (2024).
\newblock Simple controller design to achieve iso-damping robustness: Non-iterative data-driven approach based on fractional-order reference model.
\newblock \emph{arXiv preprint arXiv:2409.20375}.

\bibitem[{Zhang et~al.(2024)Zhang, Kaczmarek, and HosseinNia}]{zhang2024_SIDF_HOSIDF_ResetControl}
Zhang, X., Kaczmarek, M.B., and HosseinNia, S.H. (2024).
\newblock Frequency response analysis for reset control systems: Application to predict precision of motion systems.
\newblock \emph{Control Engineering Practice}, 152, 106063.

\end{thebibliography}
                                                   







\end{document}